# Understanding water permeation in graphene oxide membranes

Ning Wei[1], Xinsheng Peng[2], and Zhiping Xu[1,*]

[1]Applied Mechanics Laboratory, Department of Engineering Mechanics and Center for Nano and Micro Mechanics, Tsinghua University, Beijing 100084, China

[2]State Key Laboratory of Silicon Materials, Department of Materials Science and Engineering, Zhejiang University, Hangzhou 310028, China

[*]Email: xuzp@tsinghua.edu.cn

Water transport through graphene-derived membranes has gained much interest recently due to its promising potential in filtration and separation applications. In this work, we explore water permeation in graphene oxide membranes using atomistic simulations, by considering flow through interlayer gallery, expanded pores such as wrinkles of interedge spaces, and pores within the sheet. We find that although flow enhancement can be established by nanoconfinement, fast water transport through pristine graphene channels is prohibited by a prominent side-pinning effect from capillaries formed between oxidized regions. We then discuss flow enhancement in situations according to several recent experiments. These understandings are finally integrated into a complete picture to understand water permeation through the layer-by-layer and porous microstructure and could guide rational design of functional membranes for energy and environmental applications.

Graphene-derived functional membranes have recently been identified to be excellent candidate materials for filtration and separation applications because of their signature, although not well characterized, porous microstructures.[1-9] Unimpeded permeation of water through submicrometer-thick graphene oxide (GO) papers was observed with simultaneously completely impermeable performance for liquids, vapors and gases including helium.[2] The unusually high water permeability was attributed to both capillary driven force and low-friction flow confined between two-dimensional channels of pristine graphene sheets.[2] On the other hand, we reported recently that nanostrand-channelled GO ultrafiltration membranes with a network of 3-5 nm nanochannels features superior separation performance without sacrificing the rejection rate compared to unmodified GO membranes, because of their highly porous structure and significantly reduced channel length.[5] Ultrathin graphene nanofiltration membranes for water purification based on physical sieving and electrostatic interaction mechanisms were fabricated as well.[6]

In spite of the superior performance measured in experiments and its promising applications, understanding of the underlying filtration mechanism is still lacking, mainly due to the complexity in both microstructures and spatial distribution of chemical functional groups in the membrane. In this work, we elucidate the nature of water transport in GO membranes by performing atomistic simulations and continuum mechanics based analysis, which suggest a viscous nature of flow with a prominent side-pinning effect for flow between pristine graphene channels between from oxidized regions in GO sheets, and hydrogen bonds (H-bonds) between water molecules that reduces interfacial slip as well. This finding is in contradiction with the previously proposed picture of fast water flow confined between pristine channels in GO membranes where the atomistically smooth graphitic surface leads to ultralow friction. The reduced nanoconfinement effect measured directly from simulations corresponds a low flow enhancement factor of ~1-10, and the efficient water permeation through GO membranes may be attributed instead to the porous microstructures. Flow in several types of channels for water transport is then analyzed and discussed to manifest a complete picture.

**Results**

The characteristic microstructure of graphene-derived membranes is illustrated in **Figure 1**. The open path for water transport contains several types of channels that form a percolated network. We first consider a typical setup with flow across the membrane.[2, 5] In addition to cross flow from pores inside graphene sheets, interedge space between neighboring sheets, and wrinkles, the interlayer gallery may serve as a major channel due to the low edge-to-area ratio of graphene sheets that usually feature lateral sizes of micrometers. This transverse flow path introduces tortuosity in predicting overall permeability of the membrane, and could reduce efficient water transport if the channel density or connectivity is low enough. As a result, flow between GO sheets could be one of major limiting factors in this process.

Water transport between graphene sheets was reported to feature remarkable boundary slip due to the diminishing friction at the liquid-solid interface, offering remarkably high enhancement factor ~100-1000 compared to the non-slip viscous flow prediction.[10-12] However, our recent work based on molecular dynamics (MD) simulations shows that this ultrafast flow breaks down between oxidized graphene sheets due to the presence of oxygen-rich groups and their interaction with water molecules.[12] Layered water structures from monolayer, bilayers to trilayers were identified for interlayer distance $d$ below 0.7, 1.0 and 1.4 nm respectively, for both graphene and GO sheets. This value of $d$ is in the typical range for GOs immersed in water measured experimentally by Talyzin and coworkers.[13, 14] Another interesting observation of the water structure is that the distances between the first water layer (WL1) and basal plane of graphene or GO sheets are almost the same (~0.35 nm), with contrastive van der Waals and H-bond nature of the water-sheet interaction. As a result, GOs are hydrophilic with water contact angle $\theta_c$ ~30-60° and $\theta_c$ for graphene is in the range of 87-127° from experimental measurements.[15, 16] Experimental and computational studies show that the oxygen-rich groups tend to cluster on the GO sheet, forming patches of pristine and oxidized regions within the sheet,[17, 18] A percolated channel network could then be established for water transport, with a typical width of ~1-2 nm.[17, 19-21]

From all these evidences, one could expect that as a GO membrane is immersed in to the solution and water molecules enter the interlayer gallery, they start from occupying hydrophilic oxidized regions first and then open spaces between pristine parts. In recent work, capillary driven flow within these channels and low-friction monolayer water transport were proposed to explain the unimpeded permeation of water through GO membranes. The oxidized regions are considered to be spacers that separate graphene oxides with a certain interlayer distance for the capillary mechanism to work.[2] These arguments were made by assuming that water transport with the pristine channels still features significant enhancement as flow between graphene sheets. However, in this work, we find in contrast that although the water-graphene interface features low friction, the reduced water transport between oxidized regions around this pristine narrow 2D channel breaks down the expected flow enhancement. To show this effect, we performed MD simulations for model graphene and GO sheets (see **Methods** for details). In contrast to flow between graphene sheets (**Figure 2a**), the model for GO consists of a straight, pristine graphene channel with width $w_G$ = 1 nm, sandwiched between neighboring oxidized region (**Figure 2b**).[2, 18] Recent XPS measurements suggest that the oxygen-to-carbon ratio is ~40% for GO and even lower for reduced graphene oxide (RGO), and the content ratio between GO and graphene is ~1%.[22] We thus choose a typical value of number density ratio between hydroxyl groups and carbon atoms $c$ = 20% in this study.[12] The interlayer distance $d$ is 1 nm for both models and the water density profile plotted along the thickness direction $z$ in **Figure 2c** are coincident, indicating the same structure of water when confined between pristine and oxidized regions, and as a result, identical fluid properties such as density and viscosity.

We then drive water flow by applying a pressure gradient along the channel directions (see **Methods**). By plotting the velocity profiles across the channels (**Figure 3**), we find that for flow between graphene, the low friction coefficient $\lambda$ = 1.3×10$^4$ Nsm$^{-3}$ at the water-graphene interface leads to a profile with very low curvature along the thickness direction $z$, and the velocity variance is only less than 1% of its maximum amplitude. The slip length at this interface is calculated to be $L_s$ = 48.13±4.02 nm from the Green-Kubo formula.[12] However, for water flow in pristine channels of GO membranes, although a notable boundary slip still occurs between water and oxidized graphene (along the $z$ direction, as shown in **Figure 3b**), contact with neighboring oxidized regions reduces flow speed at the interfaces (**Figure 3d**). This side-pining effect leads to a parabolic profile within the pristine region along $x$, in stark contrast to the flat one measured for flow between graphene sheets (**Figure 3a** and **c**). The reduced slip length $L_s$ at this boundary is estimated to be as low as 0.9±0.35 nm (**Figure 3d**). That is to say, the presence of neighboring oxidized regions breaks down ultrafast water transport in 2D pristine graphene channels, in addition to their abovementioned role as spacers. This can be more directly seen from their three-dimensional velocity profiles in **Figure 3e** and **3f**.

**Discussion**

The well-characterized velocity profiles above allow us now to discuss flow enhancement more quantitatively measured in experiments. Considering a non-slip Poiseuille flow confined between two flat plates with distance $d$, the volumetric flux $Q$ can be estimated as

$$Q_{\text{non-slip}} = -d^3 \Delta P W / (12\eta L) \tag{1}$$

where $d$ is the channel height that should be re-defined as the hydrodynamic thickness $\delta = d - 0.5$ nm by excluding the vacuum space between water molecules and the graphene sheet, instead of the distance between basal planes of graphene or GO sheets. $\Delta P$ is the pressure difference across the channel with length $L$ and width $W$, including both pristine and oxidized regions for GO. $\eta$ is the apparent viscosity that could differ from bulk viscosity of water due to changes in the water structure under nanoconfinement.[12] Water transport in nanochannels is usually enhanced compared to $Q_{\text{non-slip}}$ because of large interfacial slip, which can be quantified by the flow enhancement factor $\varepsilon$ through the Navier slip length $L_s$ (see **Methods** for its definition) as

$$\varepsilon = Q_{\text{slip}}/Q_{\text{non-slip}} = 1 + 6L_s/\delta \tag{2}$$

The value of $L_s$ for graphene and GO with evenly distributed oxide group ($c = 20\%$) are calculated to be 138 and 0.2 nm from our MD simulations under a pressure gradient of 2.15 MPa/nm, corresponding to $\varepsilon_G = 1657$ and $\varepsilon_O = 3.7$. These values of $\varepsilon$ decrease quickly with the interlayer distance. For $d = 1.7$ nm where layered water structure is not evident, $\varepsilon$ decreases to 156.7-185.1 for graphene, and 2.1-3.0 for GO, respectively.

Now we take into account for the inhomogeneous pattern of pristine and oxidized regions in GO. We first estimate the total flux through GO is composed of contributions from the pristine and oxidized regions respectively, i.e.

$$Q_{GO} = Q_G + Q_O = -\varepsilon_G \delta^3 \Delta P n_G w_G W/(12\eta L) - \varepsilon_O \delta^3 \Delta P n_O w_O W/(12\eta L) \tag{3}$$

where $Q_G$ and $Q_O$ are the flux through pristine and oxidized regions, $n_G$ ($n_O$), $w_G$ ($w_O$) and $\varepsilon_G$ ($\varepsilon_O$) are the number density of channels per width, channel width, enhancement factor of pristine (oxidized) channels, respectively. By substituting the slip lengths calculated for graphene and GO with evenly distributed oxidized groups into **Eq. (2)**, we find then from **Eq. (3)** that the enhancement factor $Q_{GO}/Q_{non\text{-}slip}$ is 830 for $w_G = w_O = 1$ nm. This value is two orders higher than the value we measured from MD simulations that is $Q_{MD}/Q_{non\text{-}slip} = 8.06$. This inconsistence clearly indicates the aforementioned side-pinning effect on breaking down fast water transport through pristine graphene channels in GO membranes.

Our analysis on MD simulation results for the GO model shown in **Figure 2b** gives enhancement factors of 11.81 and 4.31 for flow between pristine and oxidized regions, respectively. From these results for $w_G = w_O = 1$ nm, some conclusions could be made based on **Eq. (3)**. As $w_G$ increases, the enhancement factor $\varepsilon_G$ for the pristine region increases from 11.81 to 1657 for $w_G = +\infty$. As experimental and computational evidences suggest the width of pristine is usually ~1-2 nm,[2, 17, 19-21] $\varepsilon_G$ should remain in the range of 10-100. In contrast, $\varepsilon_O$ decreases slightly with $w_O$ from 4.31 for $w_O = 1$ nm to 3.7 for $w_O = +\infty$ in GO with evenly distributed oxidized groups. As a result, one can estimate water permeation by using **Eq. (3)** using these simulated enhancement factors.

The GO membrane features rich channel microstructures in addition to the interlayer gallery as explored above, such as wider channels formed at wrinkles,[1, 23] and open spaces between edges of neighboring GO sheets or platelets. These channels are much wider than the interlayer gallery, and could reach the scale of hundreds of nanometers or even micrometers. When considered in the analysis, there is additional flux by the amount of $Q_C = -\varepsilon_C \delta^3 \Delta P n_C w_C W/(12\eta L)$, where $n_C$ and $w_C$ are the density and width of these channels. $Q_C$ for wide channels could be responsible for high permeance measured in experiments, even without any flow enhancement, i.e. $\varepsilon_C = 1$. Although because of the lack of detailed microstructural information of GO membranes such as the sheet size, layer-by-layer stacking, distribution, corrugation, and others, quantitative prediction cannot be made here. Moreover, to take into account for the complex microstructures of GO membranes, one could introduce an empirical prefactor $p/\tau$ by with the porosity $p$ and tortuosity $\tau$ fitted to the measured permeability. One of the implications of such enhancement by $Q_C$ is that one can improve the permeability by creating wider nanochannels as we have recently established through a nanostrand-templating approach.[5]

Recent work by Gao et al. shows that porous structures inside the GO sheet helps to flow enhancement as well.[6] We explore this effect, by simulating flow along open graphene edges with an interedge distance of 1 and 2 nm (**Figure 4**). The results suggest that the structure and viscosity of water within the channel are close to its bulk phase values, with a much weakened nanoconfinement effect compared to water in the interlayer gallery of the same spatial constriction. Notably, there is no apparent boundary slip, and thus for pore within graphene or GO sheets, the permeance will not be significant for a size of few nanometers. The estimated ratio between flux $Q$ and the pressure gradient - $\Delta P/L$ is $2.16\times10^{-13}$ and $3.02\times10^{-12}$ L(MPa/nm)$^{-1}$h$^{-1}$ for $w_p$ = 1 and 2 nm, which is close to $7.56\times10^{-12}$ L(MPa/nm)$^{-1}$h$^{-1}$ for GO with evenly distributed oxidized groups, but significantly lower compared to $4.56\times10^{-10}$ L(MPa/nm)$^{-1}$h$^{-1}$ for graphene with the interlayer distance $d$ = 2 nm. Moreover, by considering the random distribution of pores in the graphene, the probability for a water molecule to pass through an $N$-layer membrane with pore size $a_P$ and density $n_P$ decreases with $N$ as $2(a_P n_P)^{N-1}$. As a result, this effect shortens the transport path, but is only significant for ultrathin membranes. For example, the thickness of GO membranes is 22-53 nm in Gao et al. experiments.[6] Similar argument holds for gas separation membranes where diffusive motion of gas molecules drives the transport across membranes.[8,9] Following these findings, one could further improve the permeability by using graphene sheets with smaller size for a higher density of interedge spaces, or introducing wider channels by patterning (e.g. using nanostrands for removable templates), although the structural integrity and mechanical stability should be well maintained in these engineered membranes.[24]

According to the picture of interlayer gallery flow and **Eq. (3)**, we find that the overall flow enhancement factor $\varepsilon$ is only on the order of 10 for typical length scales of the pristine and oxidized regions in GO, e.g. $\varepsilon$ = 8.06 for $w_G$ = $w_O$ = 1 nm. Although our additional MD simulations show that $\varepsilon$ increases with $w_G$ and could reach 98.18 for $w_G$ =

4 nm and $w_O = 1$ nm, the experimental evidence that $w_G$ ~1-2 nm cannot support the explanation of high permeability by these pristine channels.[19-21] Moreover, the pristine and oxidized regions in GO are rather discretely distributed[17, 19-21] even though a percolated network could form, so the straight channel picture could break down, which leads to even lower interlayer flow rate. As a result, the measurement made recently[2] reporting enhancement factors of tens to hundreds, driven by $\Delta P$ = 23 mbar (after excluding the capillary pressure ~1-10 kbar) across a channel of ~1 μm, may arise from wider interedge spaces or channels in the membrane.

Based on above discussions, we conclude here that the reported fast flow of water across graphene-derived membranes may be attributed majorly to their porous microstructures, followed by the less significant enhancement by boundary slip. The picture of ultrafast flow between pristine graphene sheets breaks down due to a side-pinning effect by water confined between oxidized regions in GO membranes. By using adjusted density and viscosity for nanoconfined water, the interfacial slip length directly calculated from MD simulations, or the enhancement factor they lead to, one could calculate the permeability from a continuum viscous picture. These understandings not only lay the ground for future design of high-performance membranes for filtration and separation applications, but can also be extended to other membranes with similar layered microstructures, such as those constructed from $MoS_2$, $WS_2$ sheets or their hybrids,[25] although detailed characterization of their microstructures is required to allow quantitative prediction.

**Methods**

**Atomic structures.** The molecular structure of graphene oxides consists of hydroxyl, epoxy, and carbonyl groups on the basal plane, defective sites and open edges, where the hydroxyl groups were reported to be able to stay in rich in the long-living quasi-equilibrium state.[22] A typical fraction of hydroxyl species relative to the amount of carbon atoms in GO is ~20%.[22] Further reduction could yield lower concentration (13.9%-15.9%) in the reduced graphene oxide (RGO).[26] For oxidized regions of GO, we constructed hydroxyl-functionalized graphene (on both sides of the sheet) with concentration $c = n_{OH}/n_C = 20\%$, where $n_{OH}$ and $n_C$ are the number densities of hydroxyl groups and carbon atoms, respectively. The distribution of hydroxyl groups was sampled randomly in the oxidized region.

**Interatomic interaction models and molecular dynamics simulations.** Classical molecular dynamics (MD) simulations were performed using the large-scale atomic/molecular massively parallel simulator (LAMMPS).[27] The all-atom optimized potentials for liquid simulations (OPLS-AA) was used for GO, which is able to capture essential many-body terms in inter-atomic interactions, i.e. bond stretching, bond angle bending, van der Waals and electrostatic interactions.[28] This potential was successfully applied in a study of the pH-dependent behavior of graphene oxide aqueous solutions and compared to experimental results.[28] Following previous studies on similar systems, the rigid SPC/E model for water molecules was used.[11, 29] SHAKE algorithm was applied for the stretching terms between oxygen and hydrogen atoms to reduce high-frequency vibrations that require shorter time steps. The interaction between water and GO includes both van der Waals and electrostatic terms. The former one was described by the 12-6 Lennard-Jones potential $4\varepsilon_{C-O}[(\sigma_{C-O}/r)^{12} - (\sigma_{C-O}/r)^6]$ between oxygen and carbon atoms with parameters $\varepsilon_{C-O}$ = 4.063 meV and $\sigma_{C-O}$ = 0.319 nm at an interatomic distance $r$.[30] Periodic boundary conditions were applied on all directions. The channel length along the flow ($x$) direction is 12.3 nm, and the width is 2.1 nm. In order to ensure that the channel is filled at 1 a.t.m., we adjusted the interlayer distance accordingly at room temperature (300 K) for specific number of atoms. The nature of flow strongly correlates with the liquid-solid interfacial properties, one of the key parameters for which is the contact angle $\theta_c$. This set of parameters predicts a contact angle of 95º in consistent with experimental measurement.[15] The van der Waals forces were truncated at 1.0 nm and the long-range Coulomb interactions were computed by using the particle-particle particle-mesh (PPPM) algorithm.[31] To validate our model for GO, we further calculated the contact angle $\theta_c$ between water and graphene sheet with $c$ up to 20%. The results show that $\theta_c$ decrease as $c$ increases (e.g. $\theta_c$ = 29.1º at $c$ = 20%). This result is consistent with recent experimental measurement 33.7º.[7] and validates the combined OPLS-AA and SPC/E approach here can yield reasonable prediction for the water/GO hybrid system.

**Water flow simulations.** The pressure-driven water flow was simulated by directly applying forces to water molecules. This non-equilibrium molecular dynamics (NEMD) approach was widely used to explore fluid flow.[11, 12, 29] All simulations here were performed after well-converged equilibration at 300 K using the Berendsen thermostat, with the temperature calculated after removing the center-of-mass velocity. The carbon atoms in GO were frozen in the dynamic simulations to maintain the planar conformation of GO sheet as stacked in a paper or thin-film form. After the flow is driven, it usually takes a few nanoseconds to reach the steady flow state, where the external driving force and friction balance. Our discussion is based on simulation data collected in the subsequent stages. The interfacial slip was also calculated according to the definition of Navier slip length $L_s = v_s/(dv/dz)|_{z=0}$, where $v_s$ is the slip velocity at the fluid-wall interface and $dv/dz$ is the tangent of velocity profile along the normal direction $z$. The exact position of the interface ($z = 0$) is defined as the averaged position of the first water layer from the basal plane of graphene or GO sheets.

**Acknowledgements**

This work was supported by the National Natural Science Foundation of China through Grant 11222217, 11002079, 21271154, Tsinghua University Initiative Scientific Research Program 2011Z02174, and the Postdoctoral Science Foundation of China 023220045. This work was performed on the Explorer 100 cluster system of Tsinghua National Laboratory for Information Science and Technology.

7. Sun, P.; Zhu, M.; Wang, K.; Zhong, M.; Wei, J.; Wu, D.; Xu, Z.; Zhu, H. *ACS Nano* **2013,** 7, (1), 428-437.

8. Kim, H. W.; Yoon, H. W.; Yoon, S.-M.; Yoo, B. M.; Ahn, B. K.; Cho, Y. H.; Shin, H. J.; Yang, H.; Paik, U.; Kwon, S.; Choi, J.-Y.; Park, H. B. *Science* **2013,** 342, (6154), 91-95.

9. Li, H.; Song, Z.; Zhang, X.; Huang, Y.; Li, S.; Mao, Y.; Ploehn, H. J.; Bao, Y.; Yu, M. *Science* **2013,** 342, (6154), 95-98.

10. Thomas, J. A.; McGaughey, A. J. H. *Phys. Rev. Lett.* **2009,** 102, (18), 184502-4.

11. Falk, K.; Sedlmeier, F.; Joly, L.; Netz, R. R.; Bocquet, L. *Nano Lett.* **2010,** 10, (10), 4067-4073.

12. Wei, N.; Peng, X.; Xu, Z. *Phys. Rev. E* **2013,** 89, (1), 012113.

13. Talyzin, A. V.; Hausmaninger, T.; You, S.; Szabo, T. *Nanoscale* **2014,** 6, (1), 272-281.

14. Talyzin, A. V.; Luzan, S. M.; Szabó, T.; Chernyshev, D.; Dmitriev, V. *Carbon* **2011,** 49, (6), 1894-1899.

15. Rafiee, J.; Mi, X.; Gullapalli, H.; Thomas, A. V.; Yavari, F.; Shi, Y.; Ajayan, P. M.; Koratkar, N. A. *Nat. Mater.* **2012,** 11, (3), 217-222.

16. Shih, C.-J.; Wang, Q. H.; Lin, S.; Park, K.-C.; Jin, Z.; Strano, M. S.; Blankschtein, D. *Phys. Rev. Lett.* **2012,** 109, (17), 176101.

17. Zhou, S.; Bongiorno, A. *Sci. Rep.* **2013,** 3, 2484.

18. Kumar, P. V.; Bardhan, N. M.; Tongay, S.; Wu, J.; Belcher, A. M.; Grossman, J. C. *Nat. Chem.* **2013,** advance online publication.

19. Wilson, N. R.; Pandey, P. A.; Beanland, R.; Young, R. J.; Kinloch, I. A.; Gong, L.; Liu, Z.; Suenaga, K.; Rourke, J. P.; York, S. J.; Sloan, J. *ACS Nano* **2009,** 3, (9), 2547-2556.

20. Pacilé, D.; Meyer, J. C.; Fraile Rodríguez, A.; Papagno, M.; Gómez-Navarro, C.; Sundaram, R. S.; Burghard, M.; Kern, K.; Carbone, C.; Kaiser, U. *Carbon* **2011,** 49, (3), 966-972.

21. Erickson, K.; Erni, R.; Lee, Z.; Alem, N.; Gannett, W.; Zettl, A. *Adv. Mater.* **2010,** 22, (40), 4467-4472.
11

**Figures and Captions**

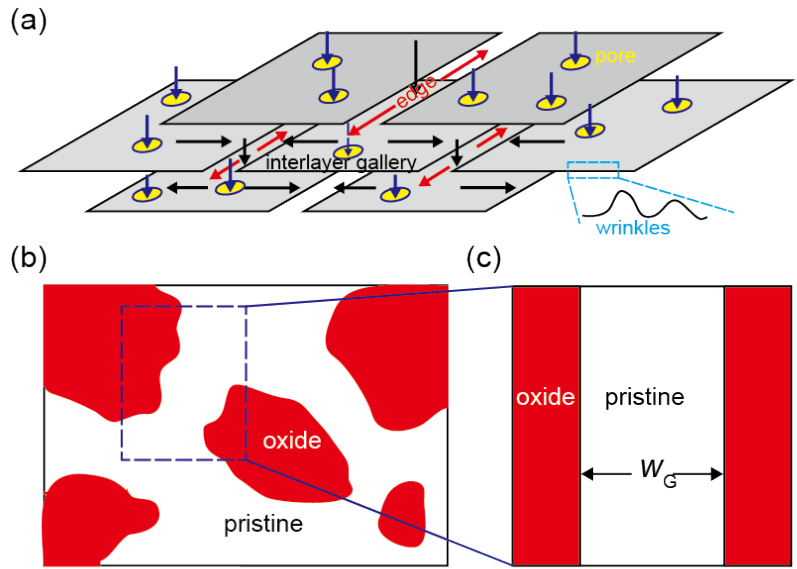

**Figure 1. Microstructures of graphene-derived membranes.** The percolated water transport channel is composed of interlayer, interedge spaces, wrinkles, and pores within the graphene sheets.

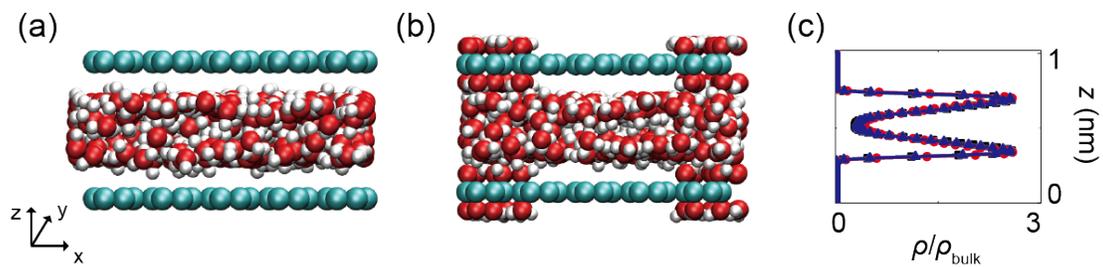

**Figure 2. Atomic structures of water molecules confined between graphene and GO sheets.** (a,b) models constructed for water flow within graphene and GO membranes. (c) shows density profiles along the thickness direction $z$, with filled squares for water between graphene sheets, dots for water between oxide regions in GO, and triangles for water between pristine regions in GO. $\rho$ and $\rho_{bulk}$ are the density of confined and bulk water respectively.

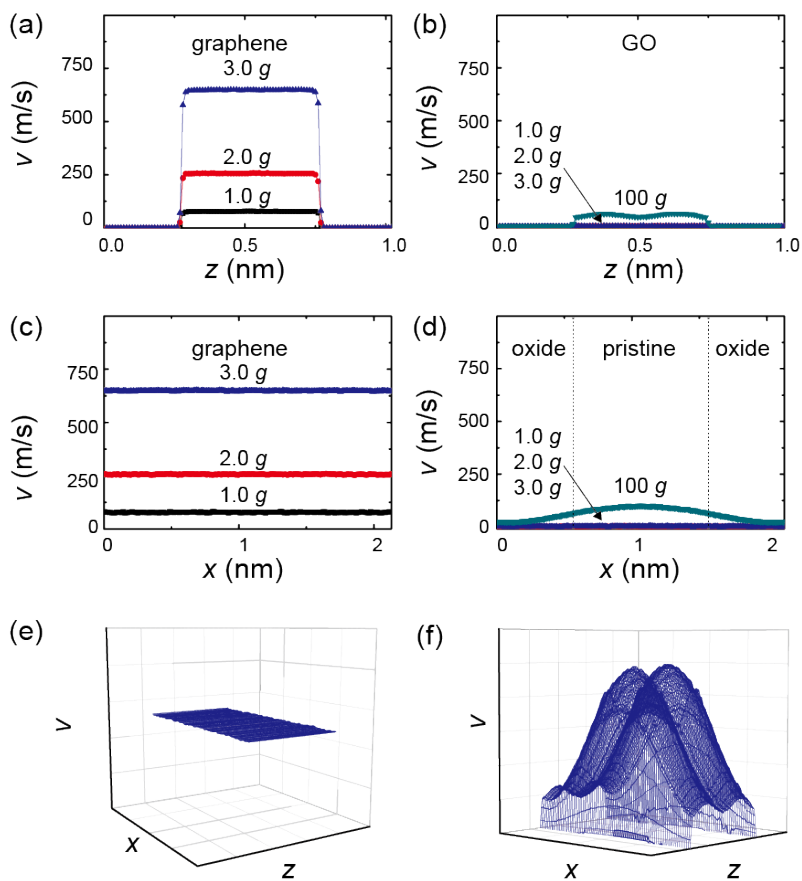

**Figure 3. Velocity profiles for water flow between graphene and GO sheets.** (a,b) for graphene and (d,e) for GO. The profiles are plotted along $x$ (a,d) and $y$ (b,e) directions, respectively. The flow profiles in three-dimensional space are also plotted in panel (c) and (f) schematically. The pressure gradient applied is $g$, $2g$ and $3g$ for data in panel (a) and (c), and $g$, $2g$, $3g$ and $100g$ for panel (b) and (d), where $g = 2.15$ MPa/nm.

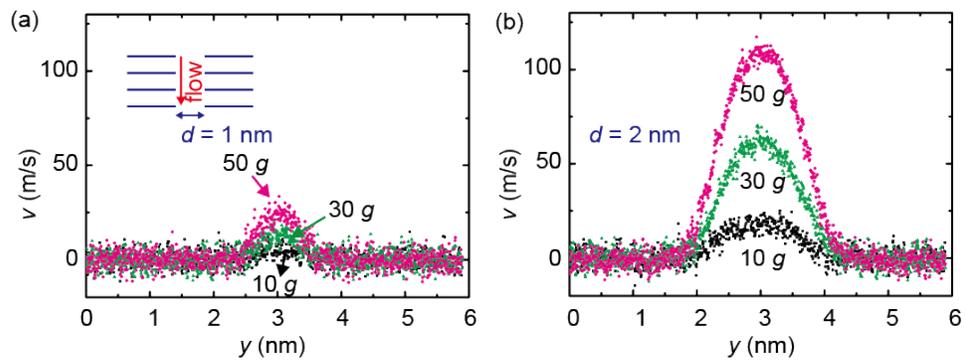

**Figure 4. Velocity profiles for water transport within channels between graphene edges.** The interedge distance is (a) 1 and (b) 2 nm, respectively. The pressure gradient applied is 10$g$, 30$g$ and 50$g$, where $g$ = 2.15 MPa/nm.